\documentclass{article}
\usepackage{ijcai21}

\usepackage{times}
\usepackage{float}
\usepackage{soul}
\usepackage{url}
\usepackage{kotex}
\usepackage[hidelinks]{hyperref}
\usepackage[utf8]{inputenc}
\usepackage[small]{caption}
\usepackage{graphicx}
\usepackage{amsmath}
\usepackage{amsthm}
\usepackage{amsfonts}
\usepackage{booktabs}
\usepackage{algorithm}
\usepackage{blindtext}
\usepackage{algorithmic}
\usepackage{subfigure}
\usepackage{mathtools,amsmath}
\usepackage{multirow}

\title{Type Anywhere You Want: An Introduction to Invisible Mobile Keyboard}

\author{
Sahng-Min Yoo$^*$
\and
Ue-Hwan Kim\footnote{Equal Contribution}\and
Yewon Hwang\And
Jong-Hwan Kim
\affiliations
KAIST, Republic of Korea\\
\emails
\{smyoo, uhkim, ywhwang, johkim\}@rit.kaist.ac.kr}
\begin{document}

\maketitle

\begin{abstract}
  Contemporary soft keyboards possess limitations: the lack of physical feedback results in an increase of typos, and the interface of soft keyboards degrades the utility of the screen. To overcome these limitations, we propose an Invisible Mobile Keyboard (IMK), which lets users freely type on the desired area without any constraints. To facilitate a data-driven IMK decoding task, we have collected the most extensive text-entry dataset (approximately 2M pairs of typing positions and the corresponding characters). Additionally, we propose our baseline decoder along with a semantic typo correction mechanism based on self-attention, which decodes such unconstrained inputs with high accuracy (96.0$\%$). Moreover, the user study reveals that the users could type faster and feel convenience and satisfaction to IMK with our decoder. Lastly, we make the source code and the dataset public to contribute to the research community.
\end{abstract}

\section{Introduction}

Text-entry plays a crucial role in Human-Computer Interaction (HCI) applications. It offers an effective and efficient way for humans to deliver messages to computers. In the early stage of text-entry research, physical keyboard-based text-entry methods were dominant. As mobile technology integrates into daily interactions in people’s lives, the need for a text entry system that suits contemporary mobile devices has emerged. In the process, soft keyboards have become the gold standard of text-entry methods for mobile devices. Soft keyboards do not require additional hardware, which improves mobility and provides an intuitive interface that guarantees usability.

However, current soft keyboards reveal some limitations: high rate of typos, lack of tactile feedback, inconsideration of users' different keyboard mental model, and large screen occupation. Since there are no clear boundaries between keys, typos commonly occur in soft keyboards by pressing an adjacent key by accident. Lack of tactile feedback requires the user's constant monitoring which could result in physical strain. Furthermore, all users are forced to type according to the given standard keyboard without any customization, although each person has a different keyboard mental model depending on their physical structure or typing habits. Lastly, with the development of Information and Communication Technology (ICT), the amount of information that smartphones can provide to humans has exploded, but soft keyboards ironically cover about half of the screens, obstructing the graphical user interface rather than delivering more information compactly.

\begin{figure}[t!]
\begin{center}
\rule{0pt}{2in}\includegraphics[width=8.3cm]{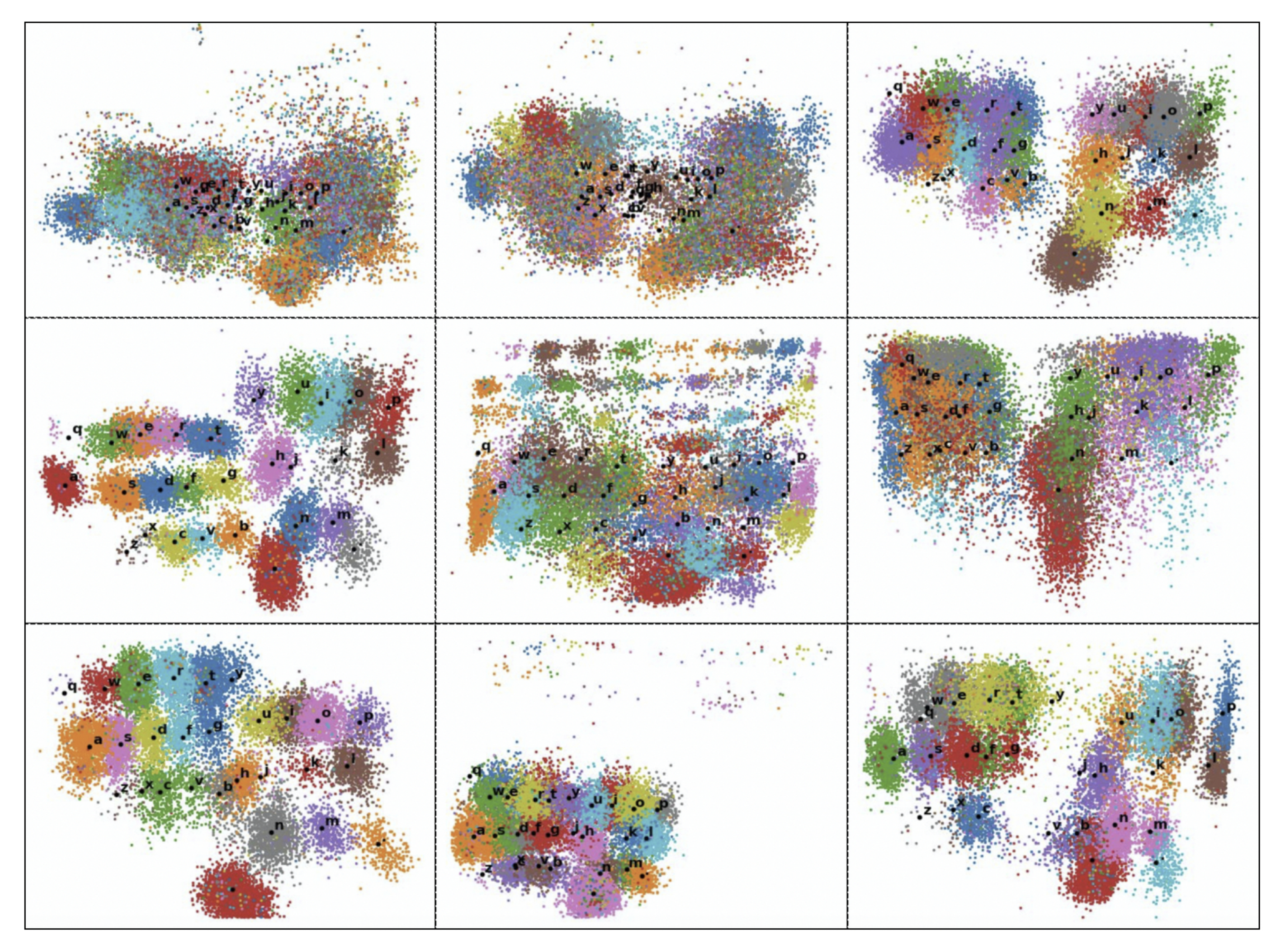}
\end{center}
   \caption{Various user mental models when typing on Invisible Mobile Keyboard. Each sub-figure is drawn with collected data of each participant.}
\label{fig:userplot}
\end{figure}

To overcome the limitations mentioned above, we propose Invisible Mobile Keyboard (IMK) with our baseline solution, Self-Attention Neural Character Decoder (SA-NCD), and collect the most extensive dataset of text-entry studies to train the proposed decoder. First of all, the proposed IMK is invisible; the entire screen can be used for displaying content which enhances the interaction between human and mobile devices; it lets users freely type text anywhere on the screen according to their typing habits. By fulfilling the IMK decoding task, users can start typing on any position at any angle as they wish.

The proposed decoder for the IMK task, SA-NCD, actualizes the full customization of soft keyboards. SA-NCD first decodes user inputs through a geometric decoder. The decoded result at this step could contain typos since users’ mental models would not precisely match the conventional keyboard layout. Then, SA-NCD semantically analyzes the user inputs (the decoded result of the geometric decoder) and corrects the typos using a language model.

In summary, the main contributions of our work are as follows:

\begin{itemize}
	\item We have collected a large-scale, richly annotated IMK dataset which contains approximately 2M pairs of touch points and text.
	\item We analyze the characteristics of user behavior on the new proposed task in-depth.
    \item We propose a novel deep neural architecture, Self-Attention Neural Character Decoder (SA-NCD) as a baseline decoder for IMK systems.
    \item We conduct a user study with the IMK prototype using our SA-NCD as a proof-of-concept to identify users' demand for research in this area.
    \item To contribute to the corresponding research community, we make the source code of SA-NCD and the collected data public. The source code and data are available at \url{https://github.com/sahngmin/InvisibleMobileKeyboard}.
\end{itemize}

\section{Related Work}
Text entry systems of mobile devices thus far use various decoding methods, where the methods can be largely divided into two categories: statistics-based and deep-learning-based methods. For statistical decoding methods, generally a probabilistic model is adopted. For instance, in \cite{findlater2012personalized}, a Naïve Bayes classifier was implemented to decode user input from an adaptive keyboard where the layout of the keyboard is altered to personalize to users. In \cite{vertanen2015velocitap}, a sentence-based decoding method was used to automatically correct typos after the entire sentence was entered. They adopted a probabilistic model for their decoder, computing the likelihood of each character using the Gaussian distribution of each key. In \cite{lu2017blindtype}, the absolute algorithm and the relative algorithm were proposed to enable eyes-free typing, where the absolute algorithm decodes user input based on the absolute position of touch endpoints, while the relative algorithm decodes based on the vectors between successive touch endpoints. Later, \cite{shi2018toast} enabled eyes-free typing on a touch screen by applying a Markov-Bayesian algorithm for input prediction, which considers the relative location between successive touch points within each hand respectively. To allow typing on an invisible keyboard on mobile devices, \cite{zhu2018typing} proposed a decoding mechanism which consists of a spatial model and a decoder. The spatial model links a touch point to a key, while the decoder predicts the next word given the history of previous words. 

Recently, with the rise of deep learning, \cite{ghosh2017neural} developed a sequence-to-sequence neural attention network model for automatic text correction and completion. This was achieved by a combination of a character-level CNN and a gated recurrent unit (GRU) \cite{chung2014empirical} encoder along with and a word-level GRU attention decoder. In addition, I-Keyboard \cite{kim2019keyboard}, another deep learning-based text entry method, was proposed. For I-Keyboard, Deep Neural Decoder (DND) was proposed as a decoding algorithm which consists of two bidirectional GRUs: one for a statistical decoding layer and the other for a language model. 
The decoding algorithm we propose in this work is deep-learning-based and employs not only a statistical model with masking process, but also a semantic model to understand the relationship between characters.

\section{Dataset}
\subsection{Dataset Construction}
To collect data, we first implemented a simple web user interface which can be accessed from any mobile device. Participants used their own mobile devices to type given phrases on the interface. The details of the data collection procedure are described in the supplementary material. Our dataset contains user initial, age, type of mobile devices, size of the screen, time taken for typing each phrase, and annotation of typed phrases with coordinate values of the typed position (x and y points). The dataset we collected is the first and only dataset for a novel IMK decoding task. Further, we have collected significantly more data compared to the datasets used in the existing text-entry research (see Table \ref{tab:ourdata}).



\begin{table}[t!]
\centering
\begin{tabular}{@{}lrrrr@{}}
\toprule
Research & $\#$ & Age        & Phrases & Points    \\ \midrule
\small{\cite{lu2017blindtype}}         & 12                       & 20$\sim$32              & 9,664                       & 40,509                     \\
\small{\cite{shi2018toast}}        & 15                       & 22 (avg.)              & 450                         & 12,669                     \\
\small{\cite{zhu2018typing}}        & 18                       & 18$\sim$50             & 2,160                       & N.A.                       \\
\small{\cite{kim2019keyboard}}        & 43   & 25 (avg.)  & 7,245   & 196,194   \\
Ours     & 88   & 18$\sim$31 & 24,400  & 1,929,079 \\ \bottomrule
\end{tabular}
\caption{Comparison of dataset sizes for text-entry research. `$\#$' and `avg.' stand for the number of participants and an average value, respectively.}
\label{tab:ourdata}
\end{table}


\begin{figure*}[!t]
\begin{center}
\includegraphics[width=0.85\linewidth]{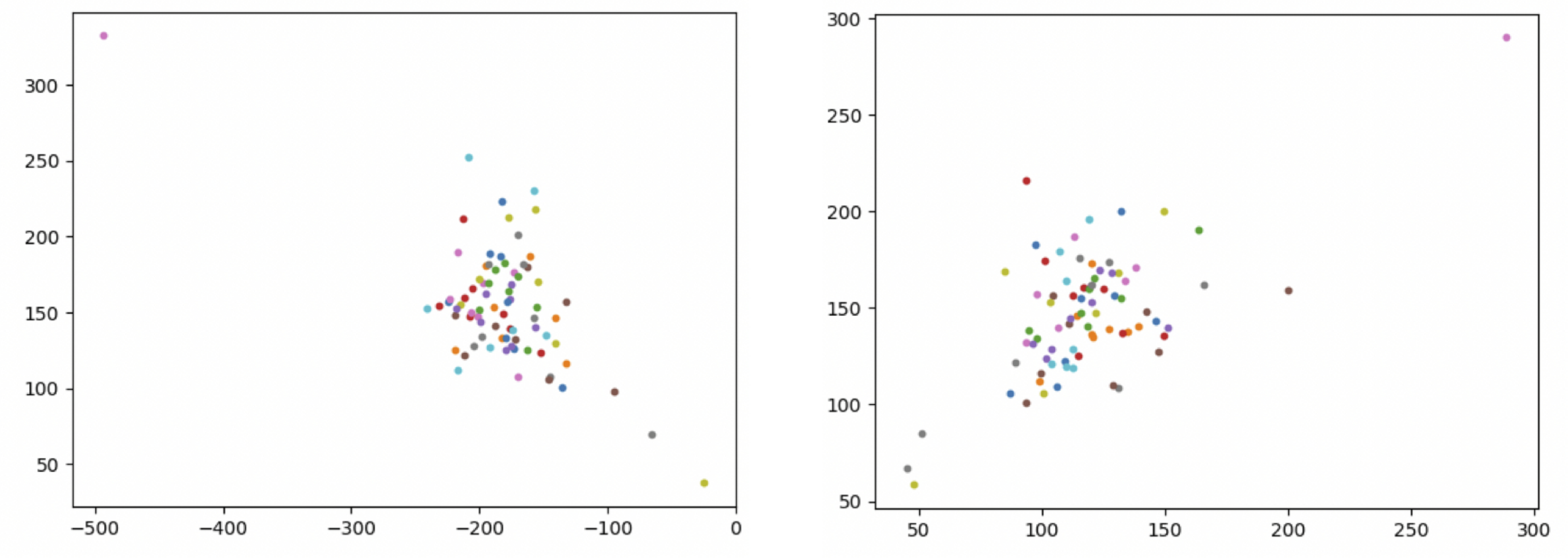}
\end{center}
   \caption{Average `q' (left) and `p' (right) positions of individual participants. Dots of the same color in left and right sub-figure are from the same person.}
\label{fig:averagepq}
\end{figure*}

\subsection{Dataset Analysis}
\label{sec:analysis}
We analyzed the user  behavior when typing on IMK and identify the characteristics of the collected data.
\subsubsection{Variance}
\begin{figure}[!t]
\begin{center}
\includegraphics[width=8.5cm]{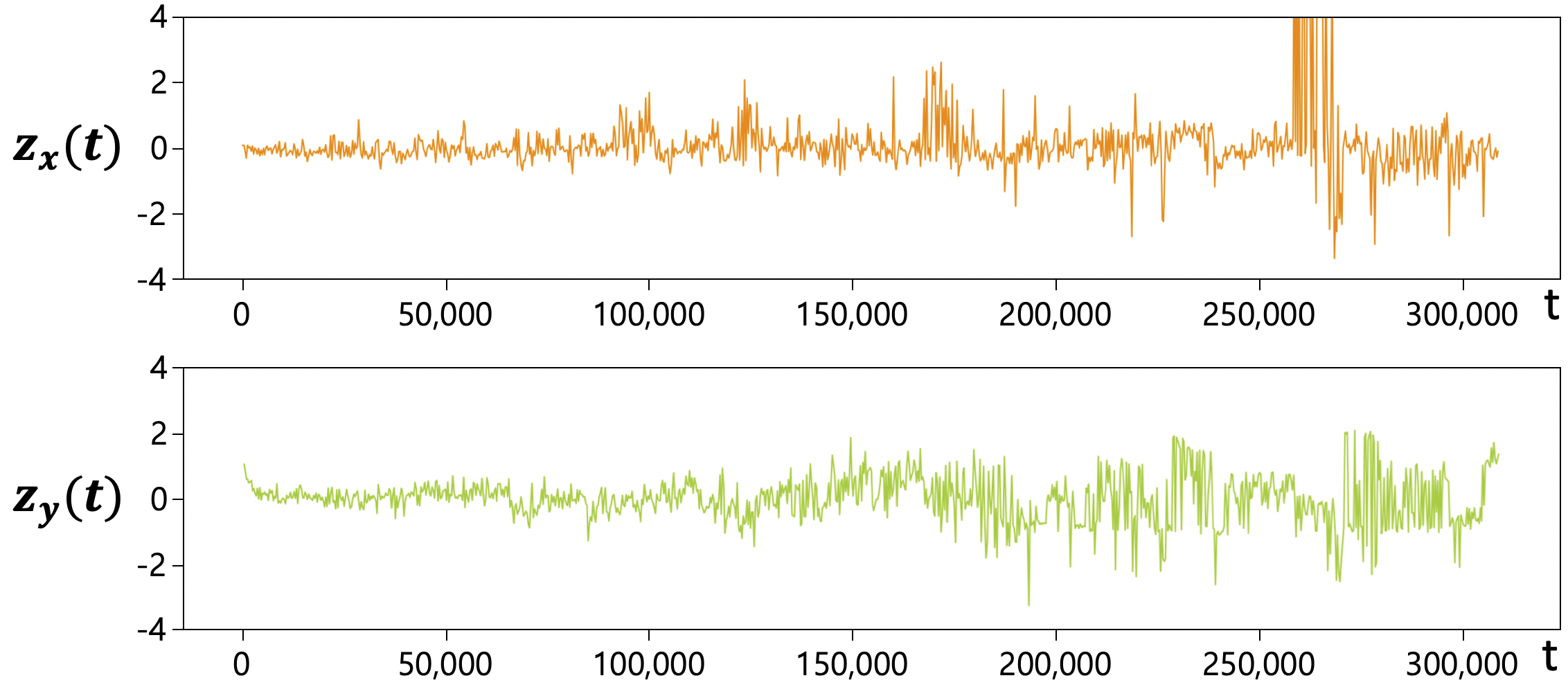}
\end{center}
   \caption{$z_x(t)$ and $z_y(t)$ value over typing time index $t$, when typing with the IMK. Values bigger than four were cut off and not drawn.}
\label{fig:overallz}
\end{figure}

We calculated the standard score (z-normalization) of x and y coordinates by averaging the entire participants and characters over typing index $t$. The standard score of x and y positions, $z_x(t)$ and $z_y(t)$, are calculated as follows:

\begin{equation}
   z_x(t) = \frac{1}{\sum_{i=1}^{N} \sum_{j=1}^v f(t, i, j)} \sum_{i=1}^{N} \sum_{j=1}^v \frac{x_{i, j}^t - \overline{x_{i, j}}}{\sigma_{i, j}} ,
    \label{eq:timez}
\end{equation}
with
\begin{displaymath}
 f(t, i, j)=\begin{cases}
    1, & \text{if } {t\leq n_{i, j}}\\
    0, & \text{otherwise}
    ,
    \end{cases}
\end{displaymath}
where $v$ is the vocabulary size, $N$ is the number of participants, $x_{i,j}^t$ is the x coordinate when $i$-th participant typing $j$-th character for $t$-th time, $\overline{x_{i, j}}$ is the mean value of $x_{i,j}^t$ taken for all $t$, and $n_{i,j}$ is the maximum number of times the $j$-th character appears for the $i$-th participant. $z_y(t)$ is calculated in the same way as $z_x(t)$, but with y coordinates. Both $z_x(t)$ and $z_y(t)$ fluctuate and the fluctuations get larger over time in both vertical and horizontal directions (see Figure \ref{fig:overallz}).

\subsubsection{Physical Dimensions}
We examined the physical dimensions of user mental models to analyze how widely and actively users moved their fingers to type on IMK. First, we analyzed the positions of ‘q’ and ‘p’ since they are the farthest keys from the space bar. The average positions of `q' and `p' for each user were inconsistent and varied significantly (see Figure \ref{fig:averagepq}). Some users typed `q' and`p' at an average of only 50 pixels apart on the x-axis, while others typed at a distance of 300 pixels.

\begin{table}[t!]
\centering
\begin{tabular}{@{}llrrrr@{}}
\toprule
\multicolumn{2}{l}{}        & mean  & std. dev. & min     & max    \\ \midrule
\multirow{2}{*}{scale}  & x & 0.99  & 0.27      & 0.36    & 3.12   \\
                        & y & 0.95  & 0.28      & -1.15   & 1.78   \\
\multirow{2}{*}{offset} & x & -2.00 & 25.68     & -180.88 & 221.90 \\
                        & y & -7.13 & 29.44     & -161.85 & 161.05 \\ \bottomrule
\end{tabular}
\caption{Statistics of collected data in x-axis and y-axis directions.}
\label{tab:scaleoffset}
\end{table}

In addition, we calculated statistics of the physical dimensions through the average distance between `space' and `p' in each sentence. Since `p' appeared 23.9 times more than `q' in our dataset, `p' was adopted for meaningful statistical analysis. We omitted the sentences that did not include `p’ for the analysis. Moreover, we use the average positions of `space' and `p' for the sentences where `space' and `p' appeared more than once. Table \ref{tab:scaleoffset} summarizes the mean, standard deviation(std. dev.), min and max values of the scale, and offset of user mental models.

The first distance between `space' and `p' was defined as scale 1, and the offset was calculated by taking the difference between the average position of the first 10 inputs and the last 10 inputs. From the statistics, it can be seen that the users type on IMK with their own mental keyboards of inconsistent scales, and the offset over time is large. The fact that the minimum of the y-axis scale is negative means that data also includes some typos such that `p' key is entered below the space bar.

As the analysis suggests, decoding our IMK dataset is a challenging task. Additionally, as Figure \ref{fig:userplot} shows, characters are not clustered neatly and typing patterns differ from user to user. Plus, the typing patterns of the right and left hands were distinguishable. Lastly, the distribution of the data consistently changed over time (see Figure \ref{fig:overallz}) and the scale of the keyboard size was undefinable (see Figure \ref{fig:averagepq}).

\begin{figure*}[!t]
\centering
\begin{center}
\includegraphics[width=0.95\linewidth]{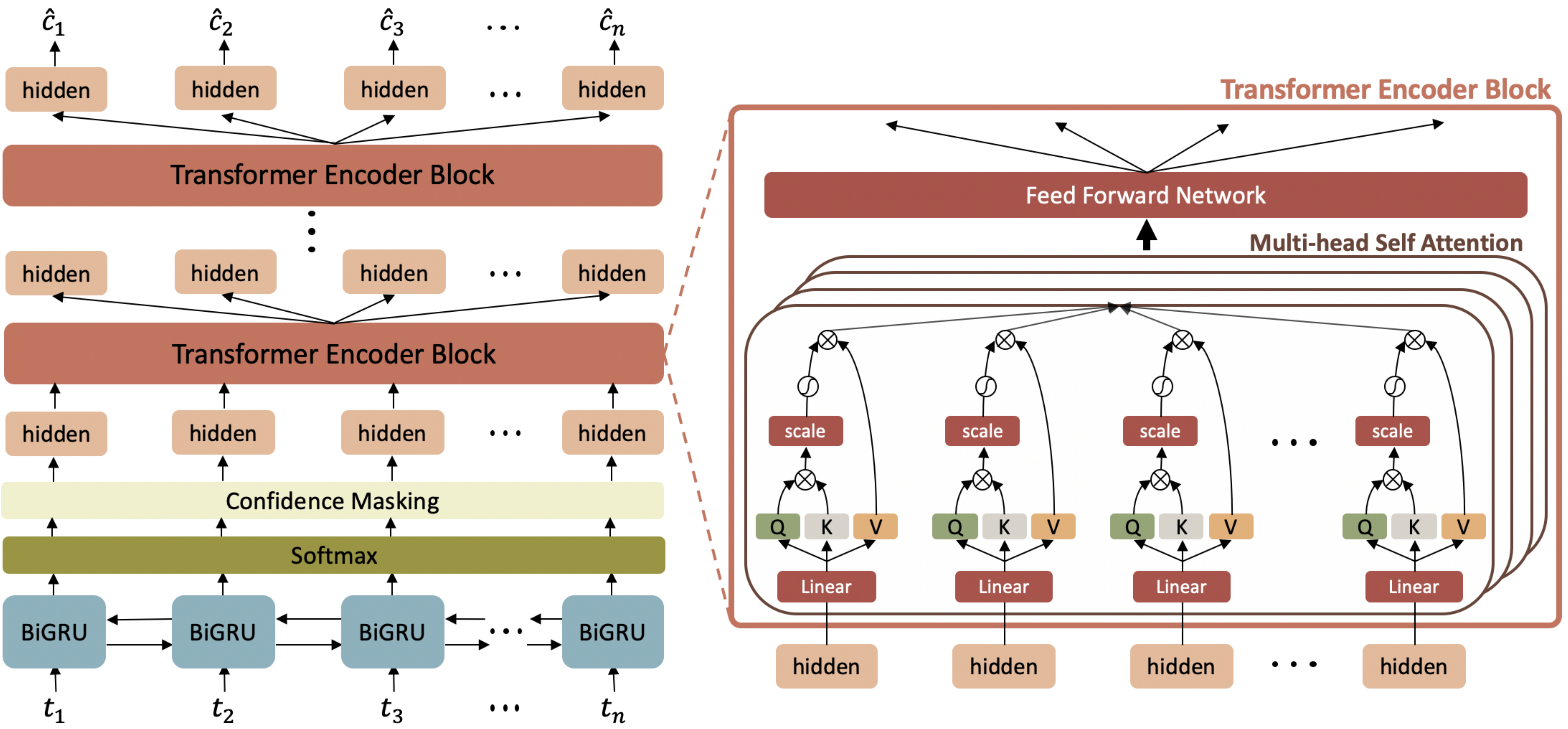}
\end{center}
   \caption{Network architecture of Self-Attention Neural Character Decoder. Q, K, and V stand for Query, Key, and Value of self-attention from Transformer.}
\label{fig:SANCD}
\end{figure*}

\section{A Baseline Approach for IMK Decoder}
\subsection{Problem Formulation}
We propose a novel baseline approach for the IMK decoding task. Specifically, taking a sequence of coordinate values as input, we aim to decode a phrase which users intended to type on an invisible layout.
Let $\hat{\mathbb{C}} = \{\hat{c_{1}}, ..., \hat{c_{n}}\}$ denote a predicted character sequence by the decoding algorithm. The decoder aims to find the character sequence, $\hat{\mathbb{C}}$, with the highest conditional probability given a typing input, $T = \{t_1, t_2, ..., t_n\}$, as follows:

\begin{equation}
    \hat{\mathbb{C}} = \mathop {\arg \max }\limits_{\hat{\mathbb{C}}} {P(\hat{c_{1}}, ..., \hat{c_{n}}|t_{1}, ..., t_{n})}.
    \label{eq:decoding_goal}
\end{equation}

\subsection{Network Architecture}
Our baseline approach, Self-Attention Neural Character Decoder (SA-NCD), consists of two decoder modules: 1) a geometric decoder ($G$) with its parameters $\phi_{G}$ and 2) a semantic decoder ($S$) with its parameters $\phi_{S}$. For an input sequence $X = \{x_1, x_2, ..., x_n\}$, let $G(X;\phi_{G})$ and $S(X;\phi_{S})$ denote the output from the geometric decoder and the semantic decoder, respectively. The geometric decoder takes in a touch input sequence and converts it into a character sequence by using the touch locations in the invisible keyboard layout. Then, the semantic decoder corrects decoding errors in the character sequence estimated by the geometric decoder while considering semantic meanings. Here, the confidence masking process determines the input of the semantic decoder by finding locations of the errors produced by the geometric decoder. The errors could be simply a case of incorrect predictions by the geometric decoder, or a case of absolute typos made by a user.
One thing to note is that SA-NCD decodes the entire sequence input up to that point (not merely the most recent touch input). Therefore, the characters decoded in the previous steps are not fixed but can be changed by reflecting the semantic dependence.

\subsubsection{Geometric Decoder}
We use Bidirectional GRU (BiGRU) for the geometric decoder. To overcome the vanishing gradient problem, we adopt GRU as a recurrent neural network. 
In our task, even if a user enters the same character index, the input (the touch points) is different every time.
Since the inputs are not a definite character index, the recurrent network suffers from the instability accumulation if it only proceeds in uni-direction. Thus, we take the advantage of forward and backward passes of BiGRU to overcome this issue.

\subsubsection{Confidence Masking}
The length of $G(X;\phi_{G}) = \{o_1, o_2, ..., o_n\}$ is the same as the input length, and each element has the dimension size of the vocabulary size ($v$). 
When considering the $i$-th input  ($1 \leq i \leq n$) from a sequence, the probability (confidence) of the $k$-th character in the vocabulary, $p_i^k$, can be calculated by the softmax function as follows:

\begin{equation}
   Softmax(G(X;\phi_{G})) = \{(p_i^1, p_i^2, ..., p_i^v)\}_{i=1}^n ,
\label{eq:softmax}
\end{equation}
with
\begin{displaymath}
  p_{i}^{k}=\frac { exp(o_{i}^{k}) }{ \sum _{ j=1 }^{ v }{ exp(o_{i}^{j}) } },
\end{displaymath}
where $o_{i}^{j}$ is the value of $j$-th element of $o_i$ in $G(X;\phi_{G})$ and $v$ is the vocabulary size.

At this step, the touch input detected at an ambiguous location that is difficult to decode with only geometric information results in low confidence. We replace the positions with a confidence lower than $\tau$ with mask tokens to provide the masked character sequence as input to the semantic decoder (confidence masking process). The following $CM$ and $Mask$ refer to the confidence masking process and the individual masking function as follows:
\begin{equation}
  CM(\{x_1, ..., x_n\}) = \{Mask(x_1), ..., Mask(x_n)\},
  \label{eq:confidence_masking}
\end{equation}
with
\begin{displaymath}
  Mask(x_i)=\begin{cases}
    Embed( \mathop {\arg \max }\limits_{j} {p_i^j}) , & \text{if $\max p_i^j \geq \tau $}\\
    Embed(j_{[mask]}), & \text{otherwise},
  \end{cases}
\end{displaymath}
where $x_i = (p_i^1, p_i^2, ..., p_i^v)$ and $j_{[mask]}$ is the index of mask token. $Embed$ is an embedding layer which encodes index to an embedded vector.

\subsubsection{Semantic Decoder}
We use the Transformer \cite{vaswani2017attention} encoder architecture as a semantic decoder. The embedded hidden state that has passed the geometric decoder followed by confidence masking is processed by the semantic decoder with the self-attention mechanism of Transformer. Here, the semantic decoder acts as a character language model, replacing the masked characters with appropriate characters.
Through the proposed network architecture (see Figure \ref{fig:SANCD}), the decoded character sequence $\hat{\mathbb{C}}$ in Equation (\ref{eq:decoding_goal}) can be expressed as follows:

\begin{equation}
    \hat{\mathbb{C}} = S(CM(Softmax(G(X;\phi_{G})));\phi_{S}).
\label{eq:output}
\end{equation}

\subsubsection{Training Scheme}
We pre-train both the geometric decoder and the semantic decoder to enhance their respective roles. In particular, we pre-process One Billion Word Benchmark (1BW) \cite{chelba2014one} to train the semantic decoder as a masked character language model by following the same pre-training scheme of BERT \cite{devlin2019bert}. Then, we fine-tune the two decoder modules together by first freezing $\phi_{G}$ and only training $S(X;\phi_{S})$, and then training $G(X;\phi_{G})$ while freezing $\phi_{S}$ repeatedly.

\section{Results}
\subsection{Experiment Settings}

\subsubsection{Data Split}
\label{sec:data_split}
Of the 88 participants of our study, 78 participants collected data using 300 sentences of 1BW, and each five of remaining participants collected data using 100 phrases of MacKenzie set (MacK) \cite{mackenzie2003phrase} and 1,000 most commonly used English phrases set (Common) \cite{EnglishSpeak}, respectively. Of these, we only utilized the data collected from 1BW data to train our decoding algorithm. We used the remaining MacK and Common sets for evaluation purposes to assess whether the proposed decoder performs adequately on sentences it has never seen before. 

We divided the 1BW data into a train (from 70 participants), validation (from 3 participants), and test (from 5 participants) sets. We divided the dataset by person rather than combining all the datasets and dividing it by ratio. This split method would verify if SA-NCD can decode sentences even from a new person with an unseen mental model. 

\subsubsection{Metrics}
We used CER (Character Error Rate) and WER (Word Error Rate) as evaluation metrics defined as follows:

\begin{equation}
    CER = \frac{MCD(\hat{\mathbb{C}}, P)}{length_{c}(P)} \times 100 \thinspace (\%),
\end{equation}
where $MCD(\hat{\mathbb{C}}, P)$ is the minimum character distance between the decoded phrase $\hat{\mathbb{C}}$ and the ground-truth phrase $P$ and $length_{c}(P)$ is the number of characters in $P$. Similarly, WER is defined as
\begin{equation}
    WER = \frac{MWD(\hat{\mathbb{C}}, P)}{length_{w}(P)} \times 100 \thinspace (\%),
\end{equation}
where $MWD(\hat{\mathbb{C}}, P)$ is the minimum word distance between $\hat{\mathbb{C}}$ and $P$ and $length_{w}(P)$ is the number of words in $P$. CER and WER count the number of insertions, deletions and substitutions of characters or words to transform $\hat{\mathbb{C}}$ into $P$.

\subsubsection{Compared Models}
We verified the performance of the proposed baseline SA-NCD compared to several sequential neural models. We implemented representative deep neural networks designed for sequential data: GRU, LSTM, Transformer, and I-keyboard which is the state-of-the-art method of the invisible keyboard task.

We measured the performance of various networks by varying the sizes which is denoted as ($L$, $d_h$), where $L$ is the number of stacked layers and $d_h$ is the size of the hidden state. In order to thoroughly check the performance according to shallow to deep, narrow to wide networks, we adopted 4 and 12 for $L$, and 256 and 512 for $d_h$. 

\subsubsection{Implementation Details}
We optimized $\phi_{G}$ with SGD update, with the learning rate of 3.0 and gradient clipping at 0.5. For $\phi_{S}$, we used the Adam optimizer with learning rate 1e-4. Threshold $\tau$ for confidence masking was 0.45 and the vocabulary size $v$ was 31. We minimized Cross-entropy loss to optimize the parameters. In addition, to quickly propagate information within the network, we employed the auxiliary losses for the character language model \cite{al2019character}. Moreover, we altered the position of layer normalization prior to self-attention to obtain the stability of the gradient and induce rapid convergence\cite{xiong2020layer}.

We implemented deep neural network models with the open-source library PyTorch 1.4.0 version and used CUDA version 10.0 for the one GPU (GTX 1080ti) computation on a Ubuntu 16.04 workstation. We augmented our data by randomly shifting the x and y positions of each touch point by 3 or fewer pixels with a 50$\%$ probability. The validation accuracy was measured by using the validation set at every epoch, and if the maximum accuracy was not updated for more than 3 epochs, the training was terminated by early stopping. 

\subsection{Results and Analysis}

\begin{table}[h!]
\begin{tabular}{@{}lrrrrr@{}}
\toprule
\multirow{2}{*}{\begin{tabular}[c]{@{}l@{}}Decoding\\ Algorithm\end{tabular}} & \multicolumn{2}{l}{Net. Size} & \multicolumn{1}{r}{\multirow{2}{*}{MacK}}               & \multicolumn{1}{r}{\multirow{2}{*}{Common}} & \multicolumn{1}{r}{\multirow{2}{*}{Test}} \\ 
                                                                     & L            & D\_h           & \multicolumn{1}{l}{}                                    & \multicolumn{1}{l}{}                        & \multicolumn{1}{l}{}                      \\ \midrule
\multirow{4}{*}{GRU}                                                 & 4            & 256            & \begin{tabular}[c]{@{}r@{}}16.04\\ (11.52)\end{tabular}  & \begin{tabular}[c]{@{}r@{}}21.51\\ (15.69)\end{tabular} & \begin{tabular}[c]{@{}r@{}}15.76\\ (9.77)\end{tabular}  \\
                                                                     & 4            & 512            & \begin{tabular}[c]{@{}r@{}}16.17\\ (11.84)\end{tabular} & \begin{tabular}[c]{@{}r@{}}21.96\\ (16.18)\end{tabular} & \begin{tabular}[c]{@{}r@{}}16.90\\ (10.80)\end{tabular}  \\
                                                                     & 12           & 256            & \begin{tabular}[c]{@{}r@{}}82.34\\ (73.93)\end{tabular} & \begin{tabular}[c]{@{}r@{}}82.19\\ (68.70)\end{tabular} & \begin{tabular}[c]{@{}r@{}}82.28\\ (85.86)\end{tabular} \\
                                                                     & 12           & 512             & \begin{tabular}[c]{@{}r@{}}82.65\\ (71.85)\end{tabular} & \begin{tabular}[c]{@{}r@{}}82.71\\ (67.70)\end{tabular} & \begin{tabular}[c]{@{}r@{}}82.35\\ (85.17)\end{tabular} \\ \midrule
\multirow{4}{*}{LSTM}                                                & 4            & 256            & \begin{tabular}[c]{@{}r@{}}46.10\\ (23.33)\end{tabular}  & \begin{tabular}[c]{@{}r@{}}55.07\\ (27.61)\end{tabular} & \begin{tabular}[c]{@{}r@{}}47.72\\ (22.50)\end{tabular}  \\
                                                                     & 4            & 512            & \begin{tabular}[c]{@{}r@{}}53.48\\ (21.82)\end{tabular} & \begin{tabular}[c]{@{}r@{}}58.27\\ (25.97)\end{tabular} & \begin{tabular}[c]{@{}r@{}}45.90\\ (19.41)\end{tabular}  \\
                                                                     & 12           & 256            & \begin{tabular}[c]{@{}r@{}}80.61\\ (91.90)\end{tabular}  & \begin{tabular}[c]{@{}r@{}}79.15\\ (90.80)\end{tabular}  & \begin{tabular}[c]{@{}r@{}}81.78\\ (95.62)\end{tabular} \\
                                                                     & 12           & 512            & \begin{tabular}[c]{@{}r@{}}80.73\\ (91.95)\end{tabular} & \begin{tabular}[c]{@{}r@{}}78.75\\ (90.55)\end{tabular} & \begin{tabular}[c]{@{}r@{}}81.75\\ (95.54)\end{tabular} \\ \midrule
\multirow{4}{*}{I-keyboard}                                          & 4            & 256            & \begin{tabular}[c]{@{}r@{}}15.28\\ (9.91)\end{tabular}  & \begin{tabular}[c]{@{}r@{}}18.14\\ (12.00)\end{tabular} & \begin{tabular}[c]{@{}r@{}}12.52\\ (6.83)\end{tabular}  \\
                                                                     & 4            & 512            & \begin{tabular}[c]{@{}r@{}}14.44\\ (9.81)\end{tabular}  & \begin{tabular}[c]{@{}r@{}}17.50\\ (11.70)\end{tabular}  & \begin{tabular}[c]{@{}r@{}}11.82\\ (6.60)\end{tabular}  \\
                                                                     & 12           & 256            & \begin{tabular}[c]{@{}r@{}}84.03\\ (48.18)\end{tabular} & \begin{tabular}[c]{@{}r@{}}85.02\\ (45.27)\end{tabular} & \begin{tabular}[c]{@{}r@{}}83.32\\ (72.35)\end{tabular} \\
                                                                     & 12           & 512            & \begin{tabular}[c]{@{}r@{}}83.58\\ (47.64)\end{tabular}  & \begin{tabular}[c]{@{}r@{}}84.28\\ (45.44)\end{tabular} & \begin{tabular}[c]{@{}r@{}}83.28\\ (72.80)\end{tabular} \\ \midrule
\multirow{4}{*}{Transformer}                                         & 4            & 256            & \begin{tabular}[c]{@{}r@{}}16.32\\ (11.31)\end{tabular} & \begin{tabular}[c]{@{}r@{}}22.77\\ (15.24)\end{tabular} & \begin{tabular}[c]{@{}r@{}}15.39\\ (9.11)\end{tabular}  \\
                                                                     & 4            & 512            & \begin{tabular}[c]{@{}r@{}}15.91\\ (10.93)\end{tabular} & \begin{tabular}[c]{@{}r@{}}22.28\\ (14.73)\end{tabular} & \begin{tabular}[c]{@{}r@{}}15.23\\ (8.94)\end{tabular}  \\
                                                                     & 12           & 256            & \begin{tabular}[c]{@{}r@{}}16.79\\ (11.58)\end{tabular} & \begin{tabular}[c]{@{}r@{}}24.67\\ (15.97)\end{tabular} & \begin{tabular}[c]{@{}r@{}}14.16\\ (8.69)\end{tabular}  \\
                                                                     & 12           & 512            & \begin{tabular}[c]{@{}r@{}}17.45\\ (11.98)\end{tabular} & \begin{tabular}[c]{@{}r@{}}23.97\\ (16.11)\end{tabular} & \begin{tabular}[c]{@{}r@{}}14.62\\ (8.95)\end{tabular}  \\ \midrule
\multirow{4}{*}{SA-NCD}                                              & 4            & 256            & \begin{tabular}[c]{@{}r@{}}7.83\\ (5.97)\end{tabular}   & \begin{tabular}[c]{@{}r@{}}8.73\\ (6.60)\end{tabular}   & \begin{tabular}[c]{@{}r@{}}7.99\\ (4.66)\end{tabular}   \\
                                                                     & 4            & 512            & \begin{tabular}[c]{@{}r@{}}7.68\\ (5.75)\end{tabular}   & \begin{tabular}[c]{@{}r@{}}8.23\\ (6.20)\end{tabular}   & \begin{tabular}[c]{@{}r@{}}7.41\\ (4.28)\end{tabular}   \\
                                                                     & 12           & 256            & \begin{tabular}[c]{@{}r@{}}7.79\\ (5.84)\end{tabular}   & \begin{tabular}[c]{@{}r@{}}8.65\\ (6.63)\end{tabular}   & \begin{tabular}[c]{@{}r@{}}7.79\\ (4.48)\end{tabular}   \\
                                                                     & 12           & 512            & \begin{tabular}[c]{@{}r@{}}\textbf{5.51}\\ (\textbf{4.82})\end{tabular}   & \begin{tabular}[c]{@{}r@{}}\textbf{7.71}\\ (\textbf{6.12})\end{tabular}   & \begin{tabular}[c]{@{}r@{}}\textbf{6.72}\\ (\textbf{4.00})\end{tabular}   \\ \bottomrule
\end{tabular}
\caption{CER and WER (values in parantheses) of decoding algorithms with various network size (Net. Size). Three sets used for performance evaluation are described in Section \ref{sec:data_split}}
\label{tab:main}
\end{table}

\subsubsection{Decoding performance}
Table \ref{tab:main} shows the decoding performance (CER and WER) of our proposed baseline and the compared models with various network sizes. GRU showed better performance than LSTM in our task. Dealing with the gates of GRU is more advantageous rather than filtering the previous information with the cell state in the LSTM for the sequence data with geometric information.
Since I-keyboard stacked two modules of bidirectional GRU, it performed better than a single GRU. 
Yet, I-keyboard also suffered from overfitting when stacked deeply, similar to GRU and LSTM. On the other hand, Transformer and SA-NCD could take advantage of greater complexity and lead to higher performance without any overfitting when the model becomes deeper. This was because they utilized efficient multi-head self-attention. However, we found that the Transformer was incapable of extracting meaningful information in the IMK task. On the other hand, by separating the geometric decoder and the semantic decoder, SA-NCD showed the best performance. In particular, the proposed SA-NCD showed 5.51 and 7.71 CER on the MacK set and Common set, which consisted of unseen phrases from unknown users.

\subsubsection{Ablation study}

For in-depth analysis, we conducted an ablation study of our SA-NCD network. Based on the best performing $(L, d_h) = (12, 512)$ configuration of SA-NCD, we explored how SA-NCD is affected without pre-training geometric decoder ($G$), and pre-training semantic decoder ($S$) (see Table \ref{tab:ablation}). 

SA-NCD w/o pre-training $G$ suffered significantly compared to SA-NCD w/o pre-training $S$. SA-NCD w/o pre-training $G$ experienced unstable learning due to the characteristics of our input. This suggests that pre-training geometric decoder plays a key role in our training strategy because input $T$ itself contains only geometric information. Therefore the semantic decoder converged well because a stable input $G(X;\phi_{G})$ with a fixed character index was provided by the geometric decoder. On the other hand, the decoding performance of SA-NCD improved slightly by pre-training $S$.

\begin{table}[]
\begin{tabular}{@{}llrrr@{}}
\toprule
                                                                                    & Metric & MacK  & Common & Test  \\ \midrule
\multirow{2}{*}{\begin{tabular}[c]{@{}l@{}}SA-NCD\\ w/o pre-training $G$\end{tabular}} & CER    & 79.89 & 79.19  & 81.63 \\
                                                                                    & WER    & 83.36 & 82.85  & 90.06 \\ \midrule
\multirow{2}{*}{\begin{tabular}[c]{@{}l@{}}SA-NCD\\ w/o pre-training $S$\end{tabular}} & CER    & 6.04  & 8.78   & 7.47  \\
                                                                                    & WER    & 5.19  & 7.38   & 4.75  \\ \midrule
SA-NCD                                                                              & CER    & 5.51  & 7.71   & 6.72  \\
                                                                                    & WER    & 4.82  & 6.12   & 4.00  \\ \bottomrule
\end{tabular}
\caption{(Ablation study) Decoding performance of SA-NCD with and without pre-training step.}
\label{tab:ablation}
\end{table}

\subsubsection{Qualitative Visualization Results}
\begin{figure}[t!]
\centering
\begin{tabular}{cc}
\includegraphics[height=2.7cm]{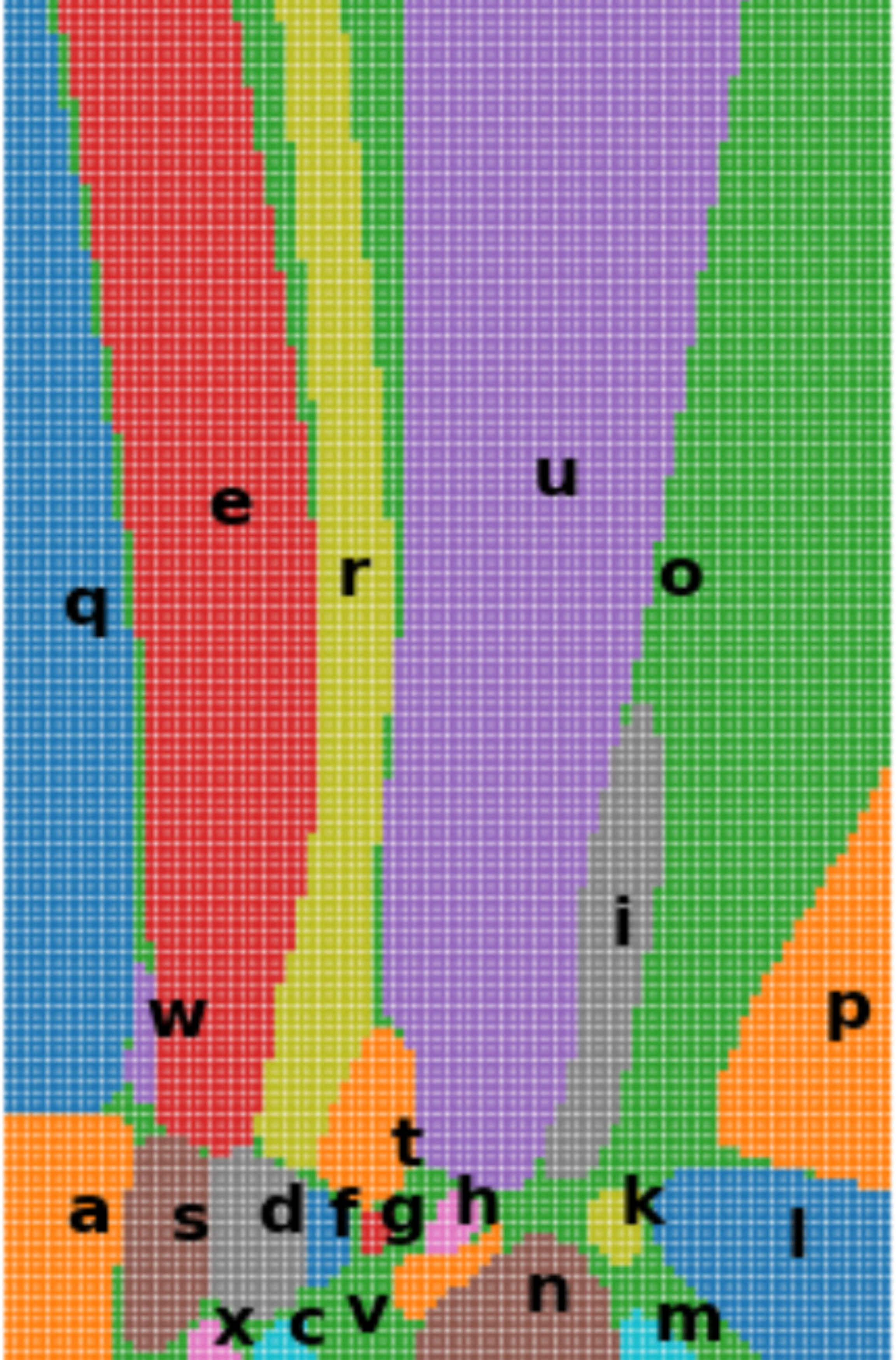}
\includegraphics[height=2.7cm]{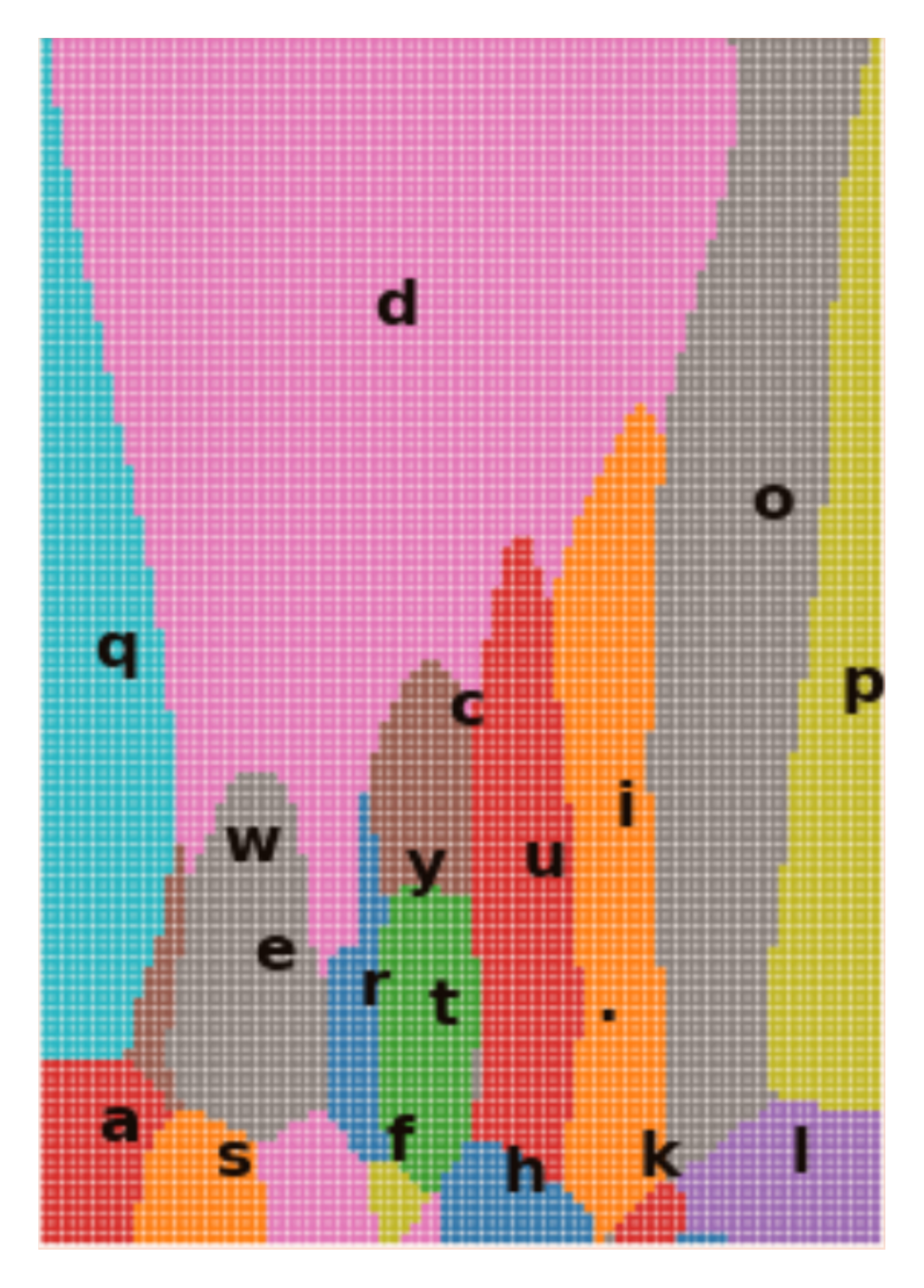}
\includegraphics[height=2.7cm]{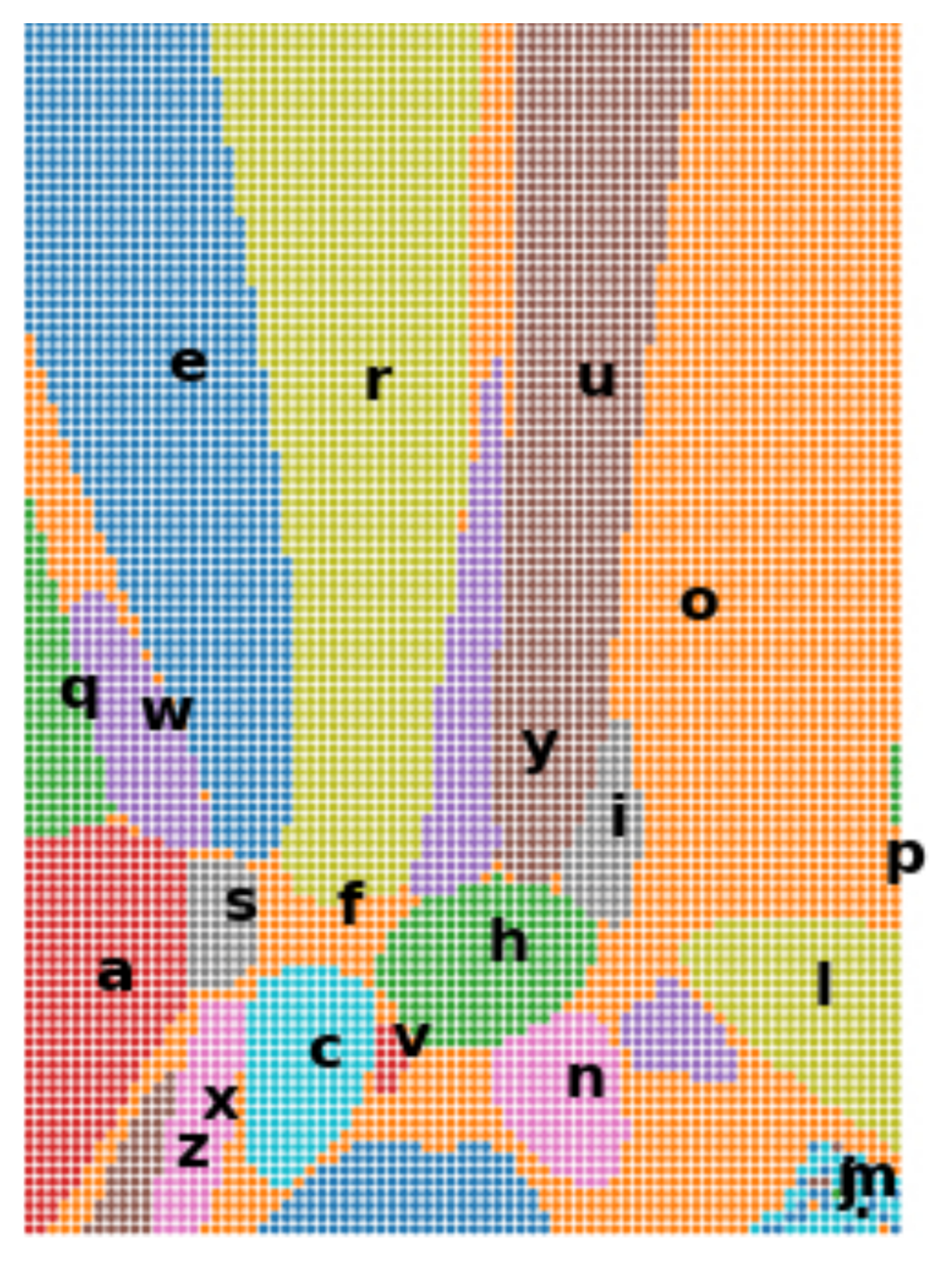}
\includegraphics[height=2.7cm]{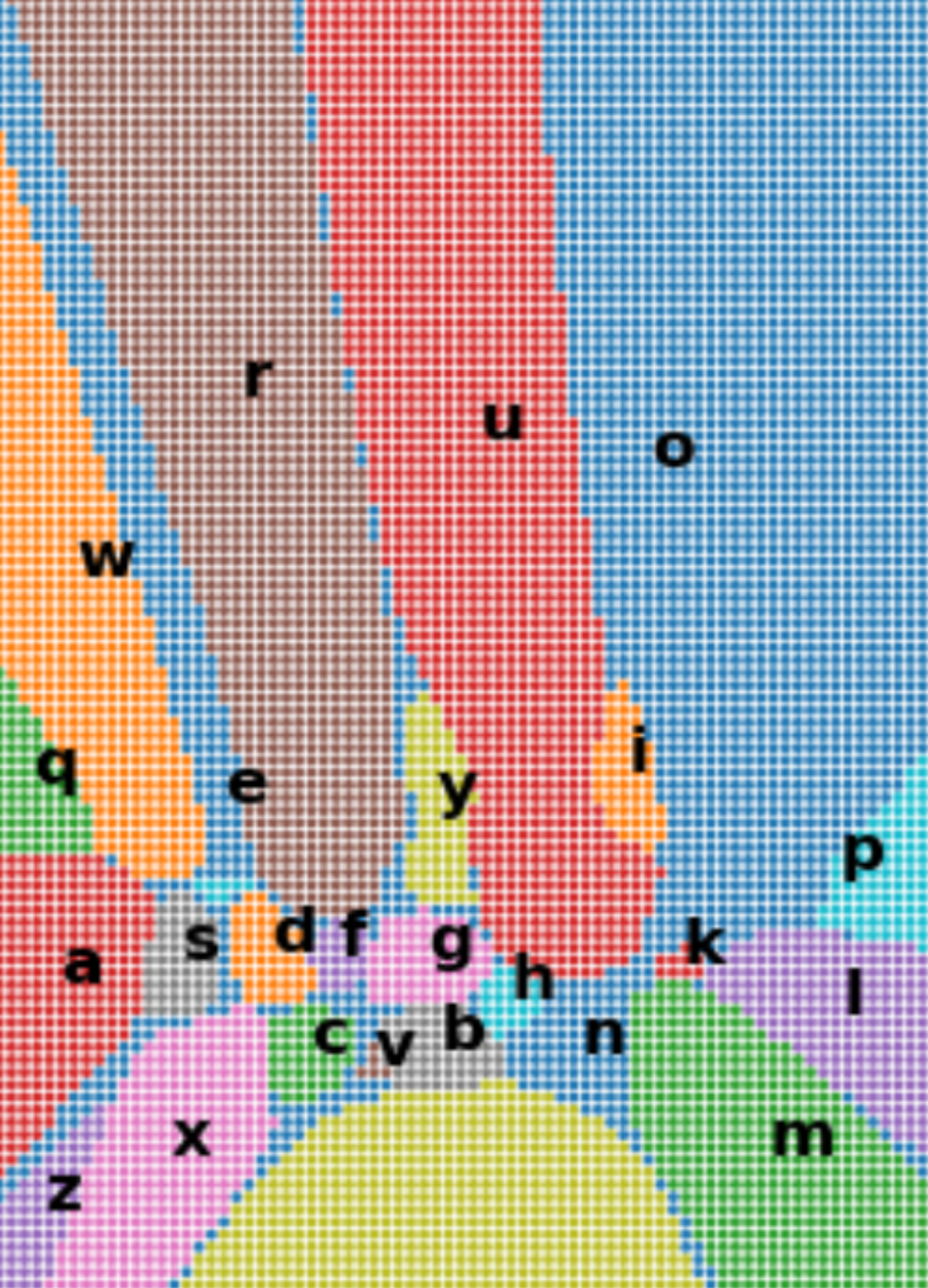}
\\(a)to stand `u' (b)at a recor`d' (c)able t`o' (d)able to`[space]'
\end{tabular}
\caption{Pixel-wise prediction for current input $t_n$ when typing each phrase. The letter inside ` ' indicates the label of the current input, $c_n$. Each bold character is marked at the average position.}
\label{fig:gridkeyboard}
\end{figure}

We visualized the outcome of the decoder at a pixel level. When the current touch input $t_n$ comes in given the previously touched input sequence $\{t_1, ..., t_{n-1}\}$, the decoder outputs the character sequence ${\hat{c}_1, ..., \hat{c}_n}$. We plotted every pair of ($t_n$, $\hat{c}_n$) to see how each predicted character by decoder is distributed over the typing area (see Figure \ref{fig:gridkeyboard}). The area corresponding to the label is located widely by the decoder.

\section{User Study}
In order to examine the usability of IMK in real-life settings, we conducted a verification experiment. The details of the user study are described in the supplementary material, including the example video.

We measured WPM (words per minute) as a metric of the typing speed of each participant, where WPM is defined as follows:
\begin{equation}
    WPM = \frac{|length_{c}(\hat{\mathbb{C}}) - 1|}{M} \times \frac{1}{n_{c}},
\end{equation}
where $M$ is the time elapsed from the first keystroke to the last keystroke for $\hat{\mathbb{C}}$ in minutes and $n_{c}$ is the average number of characters in a word.

The users were able to type 157.5$\%$ faster by using IMK than the third-party soft keyboard on their own smartphones. We can infer that typing without any strict key layouts in combination with auto typo correction by SA-NCD allowed substantial enhancement in typing speed. We further compared the speed of IMK with other types of text entry methods such as gesture-based, touch-based, and ten-finger text entry (see Table \ref{tab:wpm}). IMK surpassed all of the text entry methods, even I-keyboard which is a text entry system using ten fingers.

\begin{table}[]
\begin{tabular}{@{}llr@{}}
\toprule
Method  & Model & WPM                            \\ \midrule
Gesture & KeyScretch \cite{costagliola2011text}    & 34.5-37.4                      \\
Touch   & Invisible keyboard \cite{zhu2018typing}    & 37.9                           \\
Touch   & I-Keyboard \cite{kim2019keyboard}      & 51.3                           \\
Touch   & \textbf{Invisible Mobile Keyboard}     & \textbf{51.6} \\ \bottomrule
\end{tabular}
\caption{WPM comparison of IMK and other types of text entry methods.}
\label{tab:wpm}
\end{table}

\begin{figure}[t!]
\begin{center}
\includegraphics[width=0.45\textwidth]{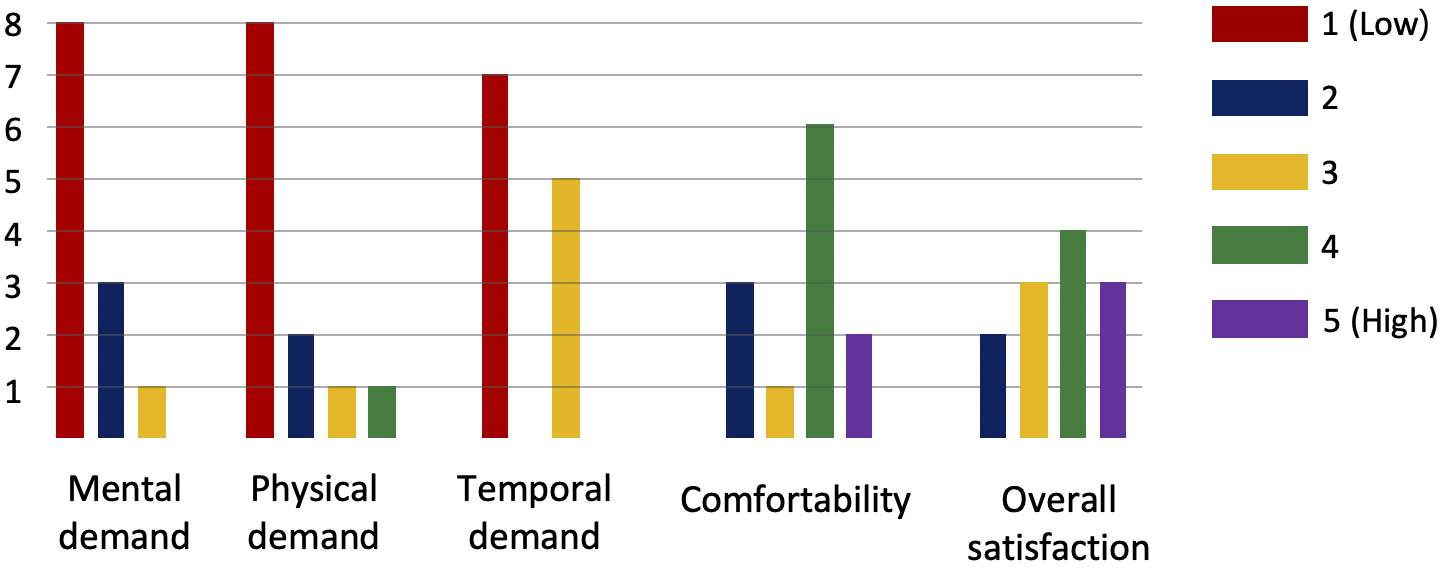}
\end{center}
   \caption{Survey results after experiencing IMK. Mental demand, physical demand, temporal demand, comfortability, and overall satisfaction were rated by each participant on a 5-Likert scale.}
\label{fig:verification}
\end{figure}

We conducted a survey after the verification experiment to evaluate the satisfaction rate of IMK. As Figure \ref{fig:verification} shows users were able to type quickly while not experiencing any physical or mental fatigue when using IMK. Overall, the users showed high satisfaction with our system. Additional usability evaluation results are described in the supplementary material.

\section{Conclusion}
In this work, we proposed a novel Invisible Mobile Keyboard (IMK) decoding task that eliminated the limitations of soft keyboards while maximizing the utility of mobile devices. We collected a suitable benchmark dataset for the task and designed a baseline approach, Self-Attention Neural Character Decoder (SA-NCD), with a new deep neural network structure and training scheme. SA-NCD led to significantly lower decoding error rates than the state-of-the-art methods of other sequential tasks. Besides, through a user study, we witness a typing speed of 51.6 WPM when using our proposed decoding algorithm, which surpassed all of the existing text-entry methods. Simultaneously, the users felt low physical, mental, and temporal demand when using our IMK with high satisfaction.
The future works shall be on a front-end design for the practical implementation of IMK.


\bibliographystyle{named}
\bibliography{ijcai21}

\end{document}